**Tetrahedrally coordinated carbonates in Earth's lower mantle**

Eglantine Boulard[1,2*†], Ding Pan[3†], Giulia Galli[3], Zhenxian Liu[4], and Wendy L. Mao[1,5]

[1] Geological and Environmental Science, Stanford University, Stanford, California 94305, U.S.A.

[2] Present address: Institut NEEL, CNRS, 25 Rue des Martyrs, 38042, Grenoble, France.

[3] Institute for Molecular Engineering, University of Chicago, Chicago, IL 60637

[4] Geophysical Laboratory, Carnegie Institution of Washington, Washington, DC 20015, USA

[5] Photon Science, SLAC National Accelerator Laboratory, Menlo Park, California 94025, U.S.A.

[†] These authors contributed equally to this work

* email: eglantine.boulard@neel.cnrs.fr

**Carbonates are the main species that bring carbon deep into our planet through subduction. They are an important rock-forming mineral group, fundamentally distinct from silicates in Earth's crust in that carbon binds to three oxygen atoms, while silicon is bonded to four oxygens. Here, we present experimental evidence that under the sufficiently high pressures and high temperatures existing in the lower mantle, ferromagnesian carbonates transform to a phase with tetrahedrally coordinated carbons. Above 80 GPa, *in situ* synchrotron infrared experiments show the unequivocal spectroscopic signature of the high-pressure phase of $(Mg,Fe)CO_3$. Using *ab-initio* calculations, we assign the new IR signature to C-O bands associated with tetrahedrally coordinated carbon with asymmetric C-O bonds. Tetrahedrally coordinated carbonates are expected to exhibit substantially**

**different reactivity than low pressure three-fold coordinated carbonates, as well as different chemical properties in the liquid state. Hence this may have significant implications on carbon reservoirs and fluxes and the global geodynamic carbon cycle.**

Ferromagnesite ((Mg,Fe)$CO_3$) minerals are likely candidates for deep-Earth carbon storage and are expected to play a key role in the deep carbon cycle[e.g. 1,2]. Therefore the properties of ferromagnesite at lower mantle conditions have been the focus of many theoretical and experimental studies[2–8]. Results of high-pressure experiments have suggested that the rhombohedral structure of magnesite (Mg$CO_3$) is stable up to 115 GPa at 2,000 – 3,000 K, and that it transforms to a new structure at higher pressures[2]. On the other hand, first-principles calculations predict that magnesite transforms into a phase containing $CO_4$ tetrahedral groups at lower pressure (~82 GPa)[5]. Boulard et al.[6] proposed that magnesite and ferromagnesite adopt the theoretically proposed structure above 80 GPa at 2300 K, based on *in situ* X-ray diffraction (XRD) and *ex situ* spectroscopy analysis on recovered samples (electron energy loss spectroscopy and scanning transmission X-ray microscopy performed at the C-K edge). However, these observations have been controversial for two reasons: (1) Only LeBail refinements were performed on the high pressure-high temperature XRD data and hence determination of the atomic positions within the unit cell was not possible; (2) Spectroscopic analyses were performed *ex situ* on recovered samples after temperature quenching and decompression to ambient conditions, where TEM analyses showed amorphization of the sample indicating the high pressure crystal structure was not preserved.

Here, we report the first unequivocal evidence of tetrahedrally coordinated carbon in high pressure carbonates, obtained by a combined experimental and theoretical study. We perform *in situ* synchrotron infrared (IR) spectroscopic studies on ferromagnesite in diamond anvil cells (DAC) and identify a unique vibrational signature present only in the high pressure phase. We perform *ab initio* calculations of the IR spectra, which allow us to assign this vibrational signature to asymmetric, $sp^3$-like C-O bonds.

**Results**

*In situ* **infrared spectroscopy characterization**

Magnesite and siderite ($FeCO_3$) form a solid solution at ambient conditions and adopt the same structure at high pressure and temperature, except that upon substitution of Mg by Fe, the volume of the unit cell decreases by about 7% [6] (V($MgCO_3$) = 351.7 Å$^3$ at 85 GPa and 2,400 ± 150 K and V($Mg_{0.25};Fe_{0.75}CO_3$) = 328.9 Å$^3$ at 80 GPa and 300 K). We note that only the Fe-bearing phase is temperature quenchable[6], an important requirement for IR measurements in the DAC. We thus chose a natural sample of ferromagnesite with a composition ($Mg_{0.25}Fe_{0.75}$)$CO_3$ which was measured with an electron microprobe. Here we refer to the high-pressure structure of ferromagnesite as post-magnesite. *In situ* IR experiments were performed at the National Synchrotron Light Source (NSLS) on the high-pressure IR beamline U2A. Two experiments were conducted: (1) IR spectra were collected upon compression and decompression of ($Mg_{0.25}Fe_{0.75}$)$CO_3$ between 0 and 54 GPa at ambient temperature. Since the sample was not heated, ($Mg_{0.25}Fe_{0.75}$)$CO_3$ did not transform into the post-magnesite structure, but only experienced an isostructural electronic spin transition around 50 GPa [9,10]. (2) In a second set of experiments, powdered $(Mg_{0.25}Fe_{0.75})CO_3$ was compressed up to 103 GPa and then transformed into the post-magnesite phase when laser-heated at ~2100 K. *In situ* XRD was used in order to monitor the transformation into the post-magnesite phase (Supplementary Fig. 1). IR spectra were recorded at the highest pressure of 103 GPa and during decompression of the post-magnesite phase back to 0 GPa at room temperature.

At low pressure, ferromagnesite is composed of repeating, alternating layers of $(CO_3)^{2-}$ radicals (anions) and layers of divalent, positively charged cations ($Fe^{2+}$ and $Mg^{2+}$) (Fig. 1A). The mid-IR spectral absorption features result primarily from fundamental internal vibrations of the C-O bonds in the carbonate radical: the out-of-plane bending ($\nu_2$), the asymmetric stretch ($\nu_3$), and in-plane bending ($\nu_4$) modes [e.g. 11]. At ambient conditions, we measured these three modes on the polycrystalline carbonate phase (symmetry group: R-3C) at 867 $cm^{-1}$; 1460 $cm^{-1}$ and 739 $cm^{-1}$ respectively (Fig. 2A). Frequency of $\nu_3$ was determined using a thinner sample (cf. supplementary Fig. 2). These frequencies are in good agreement with those reported by previous IR studies on iron-bearing carbonates [e.g. 3,11,12]. Additional modes resulting from the combination of the three fundamental ones are also present: a band at 1811 $cm^{-1}$ which stems from the combination of $\nu_1+\nu_4$; and one at 2512 $cm^{-1}$ which corresponds to the combination $2\nu_2+\nu_3$ [13]. Under compression at room temperature, no additional IR bands were observed; rather a shift to higher wave numbers was detected for all IR bands, except for the TO component of the $\nu_2$ band which exhibited a slight negative pressure shift (-0.29 $cm^{-1}$/GPa). Such a shift is in agreement with previous studies on iron-bearing carbonates [3], and has been interpreted as stemming from the increased strength of the divalent cation-oxygen bonds under compression. We observed no hysteresis upon decompression of the untransformed carbonate.

In the second set of experiments, several new IR bands were observed after transformation into the post-magnesite phase by laser heating (Fig. 2A). At very high pressure (from 103 to 81 GPa) the most intense IR bands appeared to be saturated, which made their positions difficult to determine precisely, however our experiment showed that between 103 and 43 GPa these new IR modes gradually shift to lower wavenumbers (Fig. 2B). At 0 GPa, the experimental spectrum does not correspond to a carbonate phase and displayed IR modes at 746, 836, 872, 1092, 1480, 1823, 2349 and 2530 cm$^{-1}$. We note that Boulard et al. [6] reported a redox reaction upon laser heating of $(Mg_{0.25}Fe_{0.75})CO_3$ resulting in the formation of magnetite $(Fe_3O_4)$ together with the post-magnesite phase. However, the new IR bands observed at ambient pressure after decompression cannot be assigned to magnetite [e.g. 14,15].

**First principles calculations.**

In order to interpret the IR spectrum measured for the post-magnesite phase, we carried out first principles calculations of the IR spectra of $MgCO_3$ at low and high pressure and we identified specific vibrational modes that are present only in the post-magnesite phase. Calculation were conducted for pure $MgCO_3$ instead of the solid solution, $(Mg,Fe)CO_3$, for computational simplicity. We used density functional theory (DFT), a semi local exchange-correlation functional [16], plane wave basis sets, and pseudopotentials. We first calculated the IR spectrum of magnesite (symmetry group: R-3C), the phase of $MgCO_3$ stable at ambient conditions (see Supplementary Fig. 3). We used a rhombohedral cell, with the lattice constant fixed at the experimental value of 5.675 Å and a cell angle of 48.2° [17] (Supplementary Table 1). The computed frequencies of the TO component of the $\nu_2$, $\nu_3$ and $\nu_4$ modes are 825, 1411, and 725 cm$^{-1}$ respectively. The theoretical frequencies computed for a single crystal are lower than the experimental ones by ~ 4% [11,18]. This discrepancy is likely due to the use of the semi local exchange-correlation functional Perdew-Burke-Ernzerhof (PBE). By using the hybrid functional, B3LYP, Valenzano et al. [19] found a smaller error of ~0.5% for these three bands compared to the experimental results [11,18]. The IR spectrum of the same magnesite structure computed at a pressure of 83 GPa without allowing for any phase transition (Supplementary Table 1), shows that the TO component of the $\nu_2$ mode is weakly modified (we found a modest blue shift of 5 cm$^{-1}$), the $\nu_4$ mode (TO) is blue-shifted to 913 cm$^{-1}$, and the $\nu_3$ mode (TO) is blue-shifted to 1647 cm$^{-1}$ (Supplementary Fig. 4).

Fig. 3 shows the calculated IR spectrum of the post-magnesite phase, compared with the experimental spectrum collected at ~81 GPa. We considered the structure of the post-magnesite (symmetry group: C2/m) previously measured for MgCO$_3$ at 82 GPa and high temperature (~2300 K) by Boulard et al. [6], with lattice constants $a$= 8.39 Å, $b$= 6.41 Å $c$= 6.82 Å and $\beta$= 105.49° (Supplementary Table 2). This structure is composed of tetrahedrally coordinated carbon atoms (Fig. 1B). In our computed spectrum, the phonon momentum is along the $y$-axis (crystal direction: [010]) to take into account the LO-TO splitting (IR spectra obtained by the TO component and the phonon momenta along the $z$ and $x$ directions are shown in Supplementary Fig. 5). We found three bands, at 659, 771, and 1252 cm$^{-1}$, slightly red-shifted with respect to experiments (~715, 803 and 1304 cm$^{-1}$), again most likely because of the approximation introduced by the semi-local functional used in our calculations (Supplementary Table 3). The two bands measured at 890 and 987 cm$^{-1}$ were well reproduced by our calculations; we found 908 and 987 cm$^{-1}$, respectively. Overall the theoretical and experimental spectra were in good agreement and, most importantly, they both exhibit a band at ~1300 cm$^{-1}$ which is not present in magnesite at ambient conditions or in the IR spectrum of the isotropically compressed magnesite phase (Supplementary Fig. 4). The compressed magnesite does not exhibit any extra band between 1000 and 1400 cm$^{-1}$, providing strong evidence that the experimental and theoretical band at ~1300 cm$^{-1}$ signals the presence of a structural and bonding change.

Our geometry optimization of the tetrahedrally coordinated post-magnesite phase showed that the CO$_4$ units, with nominal negative charge of 4 electrons, are asymmetric with both C and O atoms being sp$^3$-bonded, as shown in Fig. 4 that reports hybrid orbitals (Maximally Localized Wannier Functions (MLWF)[20]) obtained from linear combinations of the eigenstates of the DFT Hamiltonian used in our calculations [21]. In the primitive cell of the high pressure phase, carbon atoms have two symmetry nonequivalent positions, C1 and C2 [5]. We found that for the CO$_4$ units consisting of the C1 atoms, the two longer C-O bonds are 1.427 and 1.375 Å, and the two shorter ones are 1.256 and 1.287 Å. For the other type of CO$_4$ units (C2 atoms), the two longer C-O bonds have the same length of 1.391 Å, while the two shorter ones are 1.285 and 1.276 Å respectively. However the bonding structure of these two units is similar and represented in both cases by the hybrid orbitals shown in Fig. 4. The vibrational modes corresponding to the shorter C-O bonds are at frequencies higher than 1400 cm$^{-1}$. At 81 GPa (Fig. 2B and 3), the band at 1452 cm$^{-1}$ originates from the stretching mode of the 1.287 Å C-O bonds and the one at 1550 cm$^{-1}$ from the stretching mode of the 1.285 and 1.276 Å C-O bonds. Finally, the modes at 1631 and 1696 cm$^{-1}$ are assigned to the stretching modes of the 1.256 Å C-O bonds. We note that in the experimental spectra, the bands between 1400 and 1700 cm$^{-1}$ are saturated and individual modes cannot be identified. The vibrational mode at 1252 cm$^{-1}$ originates from the stretching mode of the 1.375 and 1.391 Å C-O bonds, with the phonon momentum along the *y* axis (Fig. 3B). The TO component of this mode is expected at 1155 cm$^{-1}$. The bands between 1100 and 1300 cm$^{-1}$ may stem from the different LO and TO components of the 1252 cm$^{-1}$ stretching mode found in the experimental spectrum. As shown in Supplementary Fig. 5, when the phonon momentum is along the *x* or *z* directions, the frequency of the stretching mode of the longer C-O bonds is lower than 1252 cm$^{-1}$. The intensities of the LO and TO components in powder transmission spectra depend on the size and shape of the powder particles. There is an apparent band at 1304 cm$^{-1}$ in the experimental spectrum suggesting that the thickness of many sample particles on the *y* direction may be much less than the wavelength of exciting radiation (7.7 μm in air)[22].

The modes detected between 1100 and 1300 cm$^{-1}$ are at lower frequencies than the asymmetric stretching mode of the planar carbonate ion $v_3$ measured at ambient conditions (1401 cm$^{-1}$ in this study). This is consistent with C-O bond lengths of 1.375~1.391 Å found in the CO$_4$ units, which are longer than that of the carbonate ion present in magnesite at ambient conditions (~1.30 Å). As expected, a longer bond length corresponds to a lower vibrational frequency. Such a trend was also found in magnesite by Grzechnik et al [23] who observed a shift of the $v_3$ mode to lower frequency by ~114 cm$^{-1}$ when decompressing the sample from 29 GPa to ambient conditions. Consistently, our calculation showed that when releasing the pressure of magnesite, from 58 to 0 GPa, the C-O bonds increases by 0.04 Å, and thus the TO component of the $v_3$ mode redshifts by 194 cm$^{-1}$.

**Discussion**

In summary, we report the first *in situ* characterization of carbon-oxygen bonds of the post-magnesite phase. We found that at high pressure, upon transformation into the post-magnesite phase, the IR spectrum of (Mg,Fe)CO$_3$ exhibits novel, unique features not present in the low pressure spectrum, which we assigned to a fundamentally different bonding configurations of carbon. Carbon bonds transforms from sp$^2$ (in the trigonal, doubly charged carbonate anions) to sp$^3$ hybridized configurations (in the tetrahedral tetra-charged carbonate anions), which we characterized using ab initio calculations within DFT. Hence our study provides the identification of a mode at 1304 cm$^{-1}$ which may be used as a fingerprint of tetrahedrally bonded carbon in high pressure mineral phases. Carbon tetrahedrally bonded to oxygen has also been proposed to be present in high-pressure phase-V of $CO_2$ [24,25]. However in this phase, tetrahedral $CO_4$ groups are symmetric with a C-O distance of 1.35Å, similar to the longest C-O distances observed here. Measured IR spectra showed the presence of a mode in the same frequency range at 1126 cm$^{-1}$, however all modes were reported to involve simultaneous stretching and bending, unlike the pure stretching mode [24] identified here.

The dramatic change we found in the carbon environment in ferromagnesite may have significant implications on carbon reservoirs and fluxes in the lower mantle and therefore on the deep carbon cycle. For example, the new bonding configuration, identified here in the high-pressure carbonate structure is dramatically different from trigonal planar group in the ambient conditions structure. This will likely influence its chemical and physical properties such as its reactivity. Moreover, one would expect dramatic changes in the behavior of carbonate melts with increases in coordination of carbon owing to the ability of $CO_4$ to form polymerizable networks while $CO_3$ trigonal groups can not[26]. At upper mantle conditions, carbonate melts differ from silicate melts. They exhibit ultra low viscosity potentially resulting in high mobilities[27]. Preliminary theoretical studies predict that carbonate melt viscosity increases at high pressures[28] which would inhibit mobility of carbonate melts in the lower mantle and might lead to the presence of deep carbon reservoirs.

**Methods:**

***In situ* high-pressure infrared spectroscopy:**

The infrared absorbance was measured in the 500 – 4000 cm$^{-1}$ range through a symmetric Mao-type DAC mounted with type-IIa diamond anvils. Measurements were conducted with a Bruker Vertex 80/v spectrometer at the side station of the high-pressure infrared beamline U2A of the National Synchrotron Light Source. The spectral resolution was 4 cm$^{-1}$ and the beam size under the microscope was approximately 20 x 20 microns. Absorption IR spectra were obtained after normalization against a background spectrum collected on an area with KBr pressure medium only (set 1) at each pressure or in the empty DAC (set 2) at ambient pressure. A natural sample of $(Mg_{0.25}Fe_{0.75})CO_3$ siderite, provided by the mineralogical collection of University of Pierre et Marie Curie, was used in this experiment. For both sets of experiments powdered $(Mg_{0.25}Fe_{0.75})CO_3$ was loaded together with a ruby ball and the pressure was determined before and after IR absorbance data collection from the shift of the ruby fluorescence line [29]. In the first set of experiments, KBr was added as a pressure transmitting medium in a DAC with 400 micron culets. One run was performed with a sufficiently thin sample in order to obtain good quality of the high intensity band *v₃* and a second run with a thicker sample in order to get good data quality of the lower intensity bands. In the second set of experiments, the sample was loaded without pressure medium in a DAC with 70 micron culets and compressed up to 103 GPa. The sample was then transformed into the post-magnesite phase by laser heating at ~2100 K at HPCAT, Argonne National Laboratory. We used *in situ* XRD in order to monitor the transformation into the post-magnesite phase.

**DFT Calculations:**

DFT calculations were performed with the Quantum ESPRESSO (QE) package [30] and Maximally Localized Wannier Functions were computed with the Qbox code (v.1.56.2, http://eslab.ucdavis.edu/software/qbox/) [31] and the algorithm of reference [21]. We used the Perdew-Burke-Ernzerhof (PBE) exchange-correlation (xc) functional [16], plane wave basis sets and ultrasoft pseudopotentials [32]. We used a kinetic energy cut-off of 40 Ry for wavefunctions and 360 Ry for charge densities. Our structural optimizations were performed by fully relaxing all atomic coordinates within the primitive cell until the force on each atom was smaller than $10^{-4}$ Ry/Bohr. We used 3×3×3 k-point mesh for the primitive cell of magnesite and 2×2×2 k-point mesh for the post-magnesite phase. We used the unit cell of the post-magnesite and atomic positions reported in Boulard et al. [6] in which the space group is P21/c after transferring the atomic positions from Oganov et al. 2008 from C2/m to P21/c. In the DFT calculations, the point group 2/m was used.

We carried out phonon calculations using density functional perturbation theory (DFPT) [33] to obtain the IR spectra. The intensity of the *m*th IR mode defined as [34]:

$$I_m = \sum_\alpha \left| \sum_{s\beta} Z^*_{\alpha\beta}(s) e_m(s,\beta) \right|^2$$

where $\alpha$ and $\beta$ are Cartesian components, $Z^*_{\alpha\beta}(s)$ is the Born effective charge tensor of the *s*th atom, and $e_m(s,\beta)$ is the normal mode eigenvector [34]. The calculated intensities were broadened by a Gaussian function with the full width at half maximum (FWHM) of 50 cm$^{-1}$.

**References:**

1. Wood, B. J., Pawley, A. & Frost, D. R. Water and carbon in the Earth's mantle. *Philos. Trans. R. Soc. London* **354,** 1495–1511 (1996).

2. Isshiki, M. *et al.* Stability of magnesite and its high-pressure form in the lowermost mantle. *Nature* **427,** 60–63 (2004).

3. Santillán, J. & Williams, Q. A high-pressure infrared and X-ray study of $FeCO_3$ and $MnCO_3$: comparison with $CaMg(CO_3)_2$-dolomite. *Phys. Earth Planet. Inter.* **143-144,** 291–304 (2004).

4. Oganov, A. R., Glass, C. W. & Ono, S. High-pressure phases of $CaCO_3$: Crystal structure prediction and experiment. *Earth Planet. Sci. Lett.* **241,** 95–103 (2006).

5. Oganov, A. R., Ono, S., Ma, Y., Glass, C. W. & Garcia, A. Novel high-pressure structures of $MgCO_3$, $CaCO_3$ and $CO_2$ and their role in Earth's lower mantle. *Earth Planet. Sci. Lett.* **273,** 38–47 (2008).

6. Boulard, E. *et al.* New host for carbon in the deep Earth. *PNAS* **108,** 5184–5187 (2011).

7. Boulard, E. *et al.* Experimental investigation of the stability of Fe-rich carbonates in the lower mantle. *J. Geophys. Res.* **117,** 1–15 (2012).

8. Tao, R., Fei, Y. & Zhang, L. Experimental determination of siderite stability at high pressure. *Am. Mineral.* **98,** 1565–1572 (2013).

9. Lavina, B. *et al.* Siderite at lower mantle conditions and the effects of the pressure-induced spin-pairing transition. *Geophys. Res. Lett.* **36,** 1–4 (2009).

10. Farfan, G. *et al.* Bonding and structural changes in siderite at high pressure. *Am. Mineral.* **97,** 1421–1426 (2012).

11. Huang, C. K. & Kerr, P. F. Infrared study of the carbonate minerals. *Am. Mineral.* **45,** 311–324 (1960).

12. Lane, M. D. & Christensen, P. R. Thermal infrared emission spectroscopy of anhydrous carbonates. *J. Geophys. Res.* **102,** 25,581–25,592 (1997).

13. Gunasekaran, S., Anbalagan, G. & Pandi, S. Raman and infrared spectra of carbonates of calcite structure. *J. Raman Spectrosc.* **37,** 892–899 (2006).

14. Chamritski, I. & Burns, G. Infrared- and Raman-active phonons of magnetite, maghemite, and hematite: a computer simulation and spectroscopic study. *J. Phys. Chem. B* **109,** 4965–8 (2005).

15. Gasparov, L. *et al.* Infrared and Raman studies of the Verwey transition in magnetite. *Phys. Rev. B* **62,** 7939–7944 (2000).

16. Perdew, J., Burke, K. & Ernzerhof, M. Generalized Gradient Approximation Made Simple. *Phys. Rev. Lett.* **77,** 3865–3868 (1996).

17. Graf, D. L. Crystallographic tables for the rhombehedral carbonates. *Am. Mineral.* **46,** 1283–1316 (1961).

18. Hellwege, K. H., Lesch, W., Plihal, M. & Schaack, G. Zwei-Phononen-Absorptionsspektren und Dispersion der Schwingungszweige in Kristallen der Kalkspatstruktur. *Zeitschrift Phys.* **232,** 61–86 (1970).

19. Valenzano, L. *et al.* Ab initio vibrational spectra and dielectric properties of carbonates: magnesite, calcite and dolomite. *Theor. Chem. Acc.* **117,** 991–1000 (2007).

20. Marzari, N., Mostofi, A. a., Yates, J. R., Souza, I. & Vanderbilt, D. Maximally localized Wannier functions: Theory and applications. *Rev. Mod. Phys.* **84,** 1419–1475 (2012).

21. Gygi, F., Fattebert, J.-L. & Schwegler, E. Computation of Maximally Localized Wannier Functions using a simultaneous diagonalization algorithm. *Comput. Phys. Commun.* **155,** 1–6 (2003).

22. Farmer, V. C. Differing effects of particle size and shape in the infrared and Raman spectra of kaolinite. *Clay Miner.* **33,** 601–604 (1998).

23. Grzechnik, A., Simon, P., Gillet, P. & McMillan, P. An infrared study of $MgCO_3$ at high pressure. *Phys. B* **262,** 67–73 (1999).

24. Santoro, M. *et al.* Partially collapsed cristobalite structure in the non molecular phase V in $CO_2$. *Proc. Natl. Acad. Sci. U. S. A.* 2–5 (2012). doi:10.1073/pnas.1118791109

25. Datchi, F., Mallick, B., Salamat, A. & Ninet, S. Structure of Polymeric Carbon Dioxide $CO_2$-V. *Phys. Rev. Lett.* **108,** 1–5 (2012).

26. Jones, A. P., Genge, M. & Carmody, L. Carbonate Melts and Carbonatites. *Rev. Mineral. Geochemistry* **75,** 289–322 (2013).

27. Kono, Y. *et al.* Ultralow viscosity of carbonate melts at high pressures. *Nat. Commun.* **5,** 5091 (2014).

28. Jones, A. P. & Oganov, A. Carbon rich melts in the Earth's deep mantle. in *Diam. Work. Bressanone* (Nestol, F.) <http://www.univie.ac.at/Mineralogie/EMU/media/NESTOLA/Jones_1.pdf> (2010)29. Mao, H. K., Xu, J. & Bell, P. M. Calibration of the Ruby Pressure Gauge to 800 kbar under quasi-hydrostatic Conditions. *J. Geophys. Res.* **91,** 4673–4676 (1986).

30. Giannozzi, P. *et al.* QUANTUM ESPRESSO: a modular and open-source software project for quantum simulations of materials. *J. Phys. Condens. Matter* **21,** 395502 (2009).

31. Gygi, F. Architecture of Qbox : A scalable molecular dynamics code. *IBM J. Res. amd Dev.* **52,** 1–8 (2008).

32. Vanderbilt, D. Soft self-consistent pseudopotentials in generalized eigenvalue formalism. *Phys. Rev. B* **41,** 7892–7895 (1990).

33. Baroni, S., de Gironcoli, S., dal Corso, A. & Giannozzi, P. Phonons and related crystal properties from density-functional perturbation theory. *Rev. Mod. Phys.* **73,** 515–562 (2001).

34. Giannozzi, P. & Baroni, S. Vibrational and dielectric properties of C60 from density-functional perturbation theory. *J. Chem. Phys.* **100,** 8537 (1994).

**Acknowledgments:**

The use of the U2A beamline was supported by COMPRES, the Consortium for Materials Properties Research in Earth Sciences under NSF Cooperative Agreement EAR 11-57758" and the DOE-NNSA (DE-FC03-03N00144, CDAC). Use of NSLS was supported by the U.S. Department of Energy (DOE), Office of Science, Office of Basic Energy Sciences (BES), under Contract No. DE-AC02-98CH10886. *In situ* laser heating and XRD were performed at Argonne National Laboratory HPCAT (Sector 16), which is supported by DOE-NNSA, DOE-BES, and NSF. APS is supported by DOE-BES, under Contract No. DE-AC02-06CH11357. We thank C.Y. Shi for help with the sample preparation. E. Boulard and D. Pan acknowledge support from the Deep Carbon Observatory. G. Galli acknowledges support from DOE/BES DE-SC0008938. W. L. Mao acknowledges support from NSF, Geophysics Grant EAR-1141929.

**Authors contributions:**

E.B designed the study. E.B. and Z.L. conducted the experiments. E.B. analyzed the data. D.P. performed the theoretical calculations. E.B. and D.P. wrote the manuscript. All authors discussed the results and commented on the manuscript.

**Additional information**

Supplementary information is available in the online version of the paper.

**Competing financial interests**

The authors declare no competing financial interests.

**Figure captions:**

**Figure 1. Ball and stick representation of the two structures.** Ferromagnesite at ambient conditions (A) and the post-magnesite phase above 80 GPa (B). Red, grey and blue spheres represent oxygen, carbon and cations (Fe or Mg) atoms, respectively.

**Figure 2: *In-situ* IR measurement at high pressure.** (A) Experimental IR spectra collected upon compression of the ferromagnesite (black lines) at 0 and 54 GPa and upon decompression of the post-magnesite phase (red lines) at 58 and 0 GPa. The region between 1900 and 2300 cm$^{-1}$ is dominated by absorption from the diamond anvils. (B) Experimental IR spectra collected upon the decompression of the post-magnesite phase from 103 to 0 GPa. The scale bars give the absorbance scale for each panel.

**Figure 3: Theoretical IR results.** (A) Calculated IR intensities (black lines) and spectrum (solid red line) of the post-magnesite phase at 82 GPa. For comparison, the experimental spectrum collected at 81 GPa is shown as a dotted line. The scale bar gives the absorbance scale. (B) The vibrational mode at 1252 cm$^{-1}$ (marked by the arrow on panel A) is identified as a unique signature of the high pressure phase.

**Figure 4: Hybrid orbitals in C-O bonds.** Three-fold (A) and four-fold (B) coordinated carbon atoms in the magnesite and post-magnesite phases, respectively. Grey and red spheres represent C and O atoms, respectively. Green spheres represent the centers of the hybrid orbitals for the $CO_3^{2-}$ anion (A), with 24 valence electrons and 12 doubly occupied hybrids; and for the $CO_4^{4-}$ anion (B) with 32 valence electrons. The white and blue colors denote the positive and negative parts of the wave function, respectively. In magnesite both C and O are sp$^2$ bonded and three distinct orbitals were identified: σ-like orbitals localized on the C-O bonds (orbital 2), π-like

bond formed by the overlap of $p_z$ orbitals on the O and C atoms (orbital 1) and lone pairs (orbitals 3) on the oxygen atoms; in post-magnesite both C and O are $sp^3$ bonded and two distinct orbitals were identified: σ-like orbitals localized on the C-O bonds (orbital 5) and lone-pair hybrids (orbitals 4) on the oxygen atoms. Hybrid orbitals were defined using maximally localized Wanier functions, built from linear combinations of single particle electronic eigenstates (see Method section).

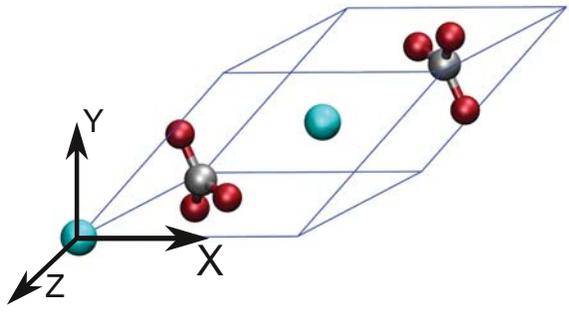 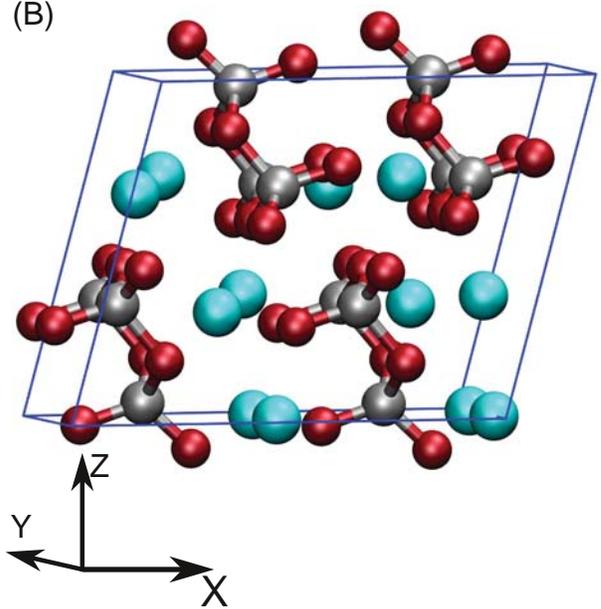

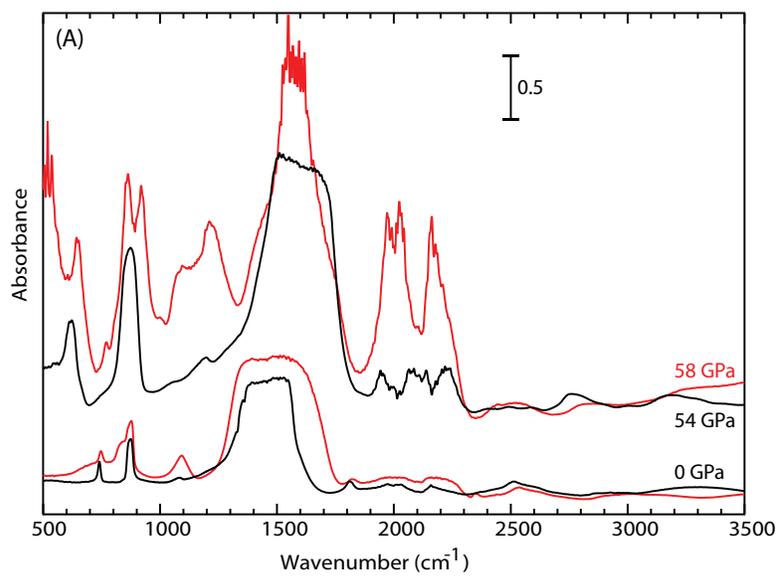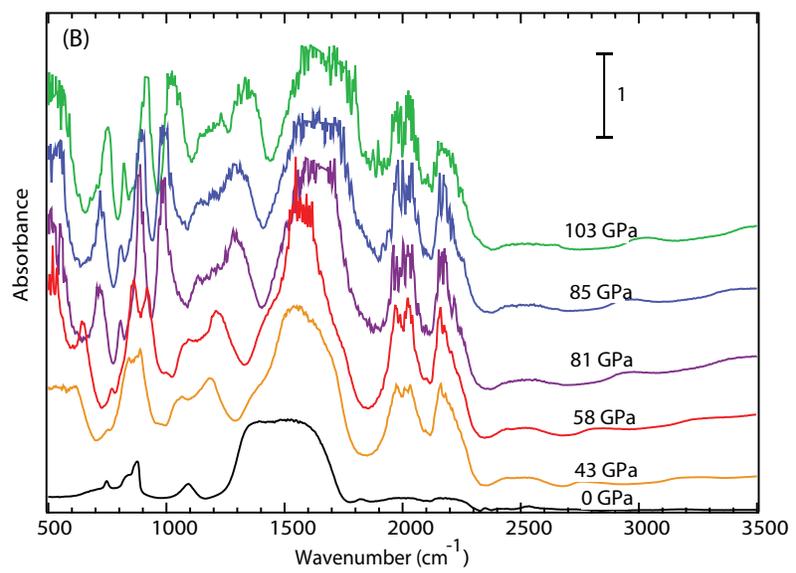

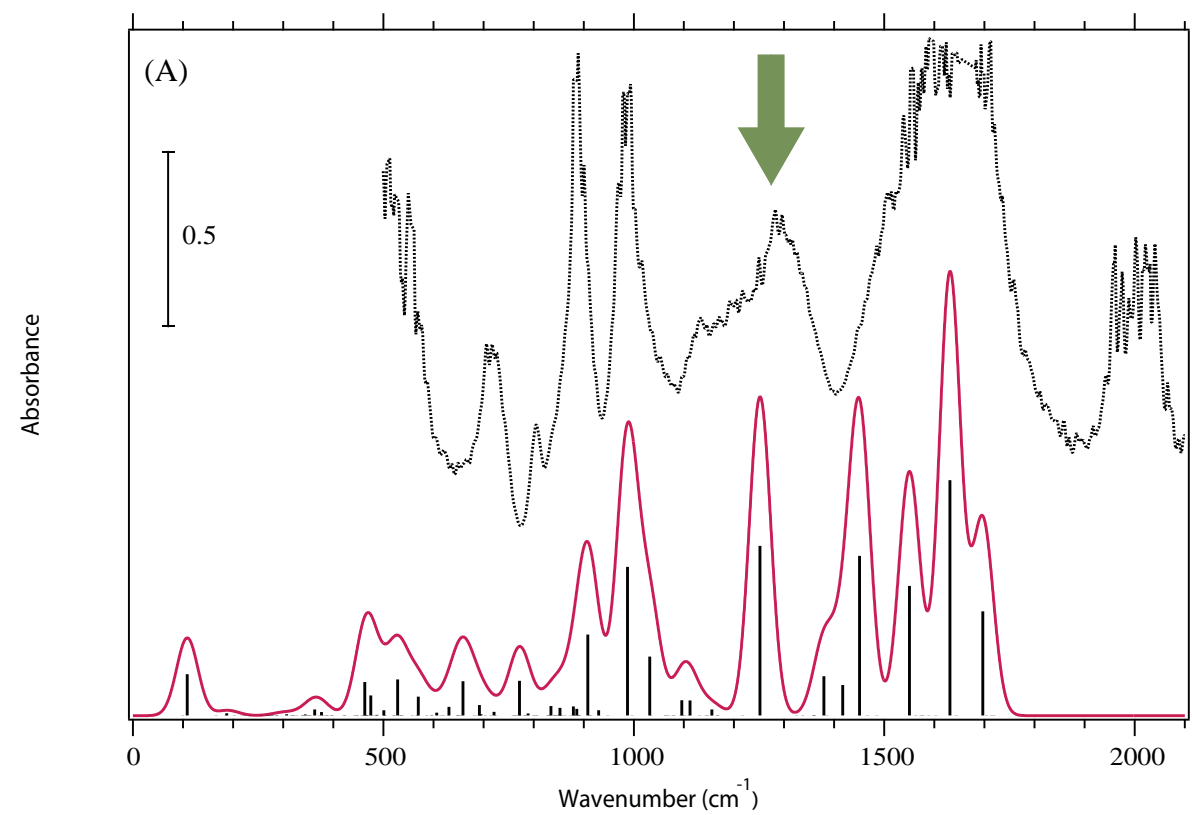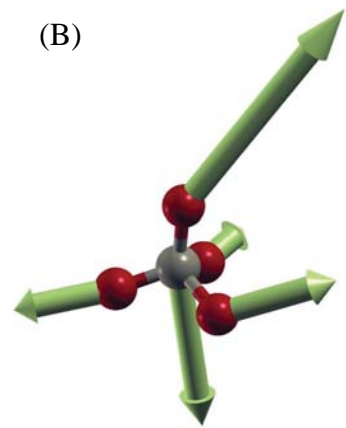

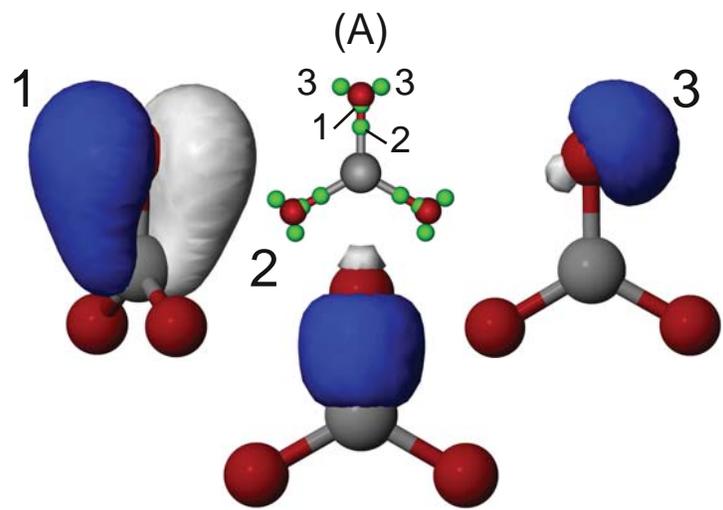
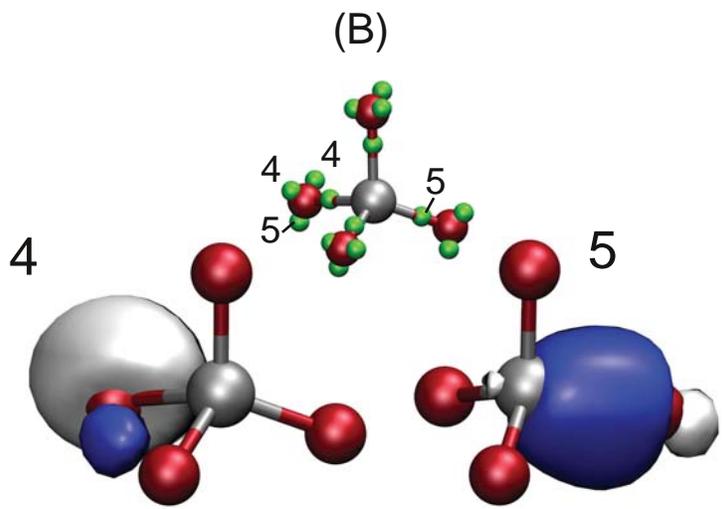

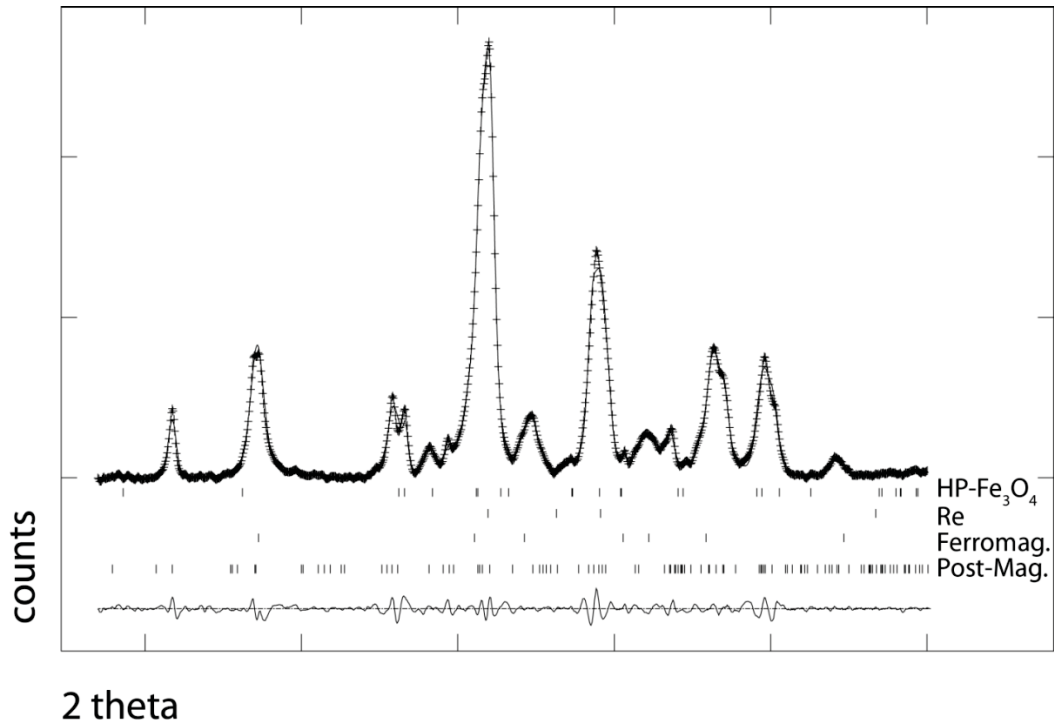

2 theta

**Supplementary Figure 1:** XRD pattern of (Mg,Fe)CO$_3$ at 103 GPa after transformation at ~2100 K (wavelength: 0.4066 Å) confirming the transformation into the post-magnesite phase. Crosses are the observed diffraction data and solid line is the profile refinement. The background has been subtracted. HP-Fe$_3$O$_4$: high-pressure phase of Fe$_3$O$_4$ described by Haavick et al. [1]; Re: Rhenium gasket; Ferromag.: ferromagnesite (spacegroup R-3c); Post-Mag: post-magnesite structure.

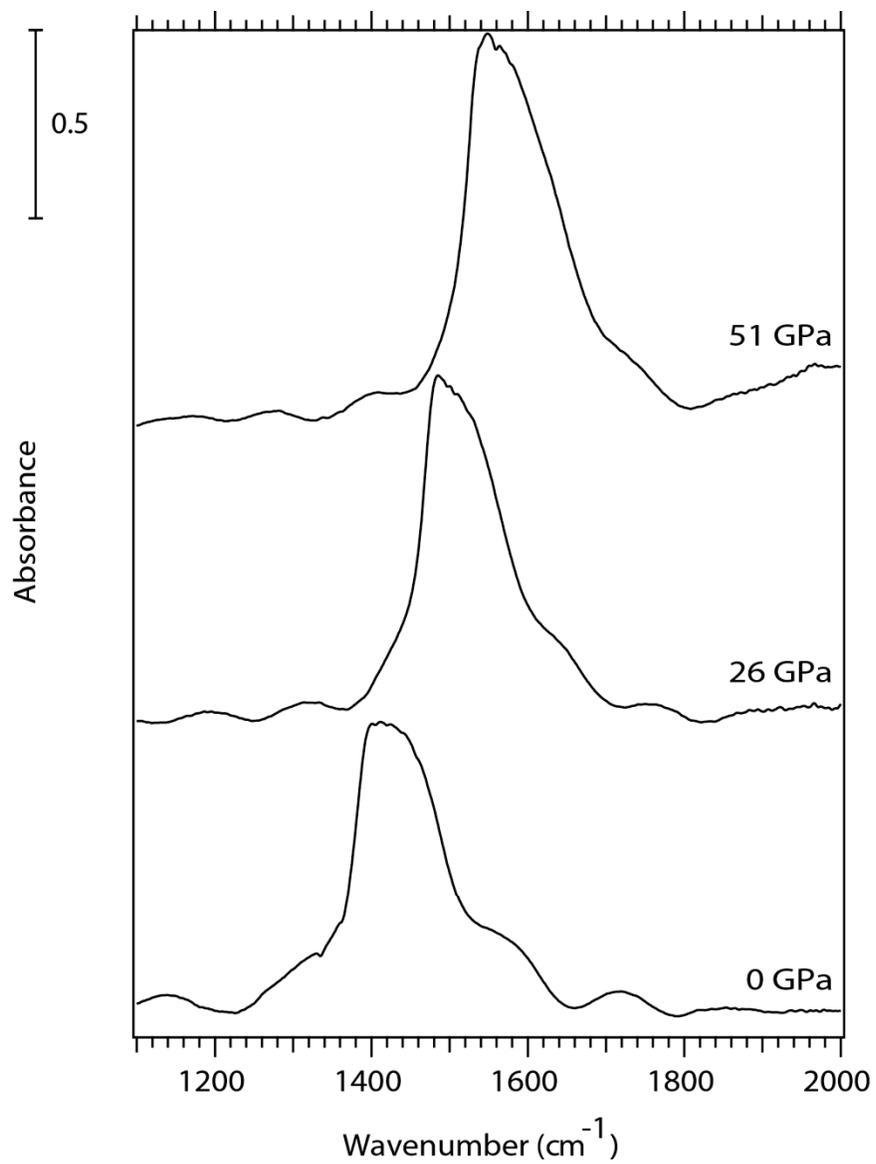

**Supplementary Figure 2:** Representative experimental IR spectra measured at upon compression on a thin sample of $(Mg_{0.25}Fe_{0.75})CO_3$ at room temperature. The scale bar gives the absorbance scale.

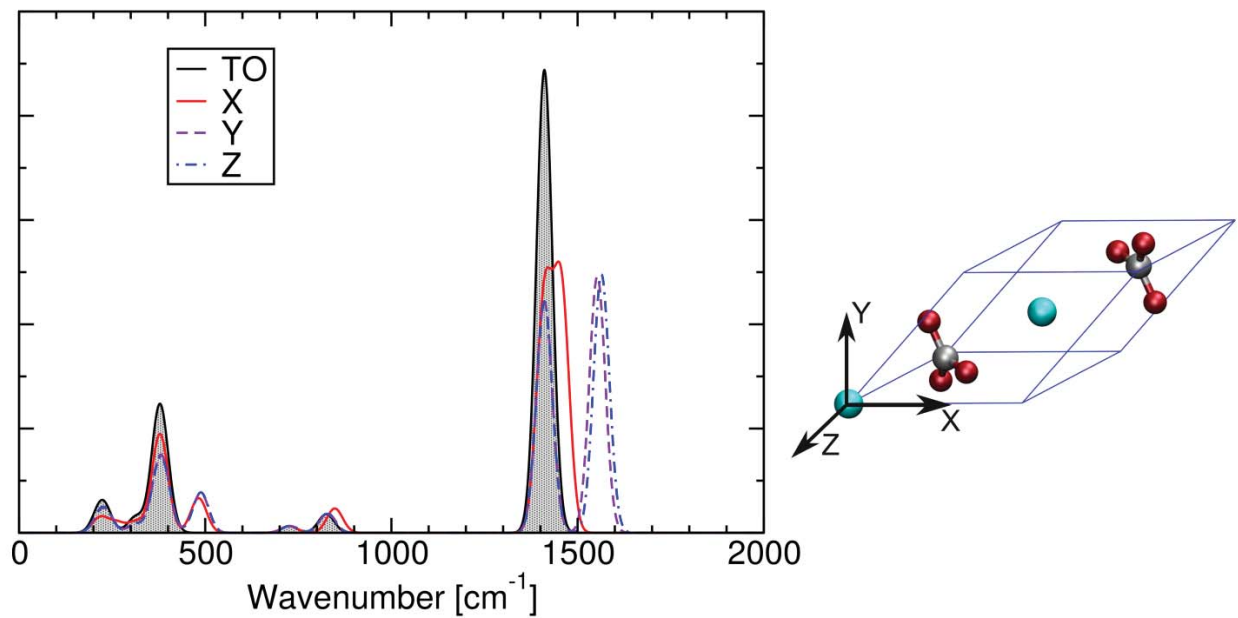

**Supplementary Figure 3:** Calculated IR spectra of magnesite at ambient condition (using experimental lattice constants). The LO-TO splitting is along *x*, *y* and *z* directions, respectively, and pure TO mode is shown by the gray shading.

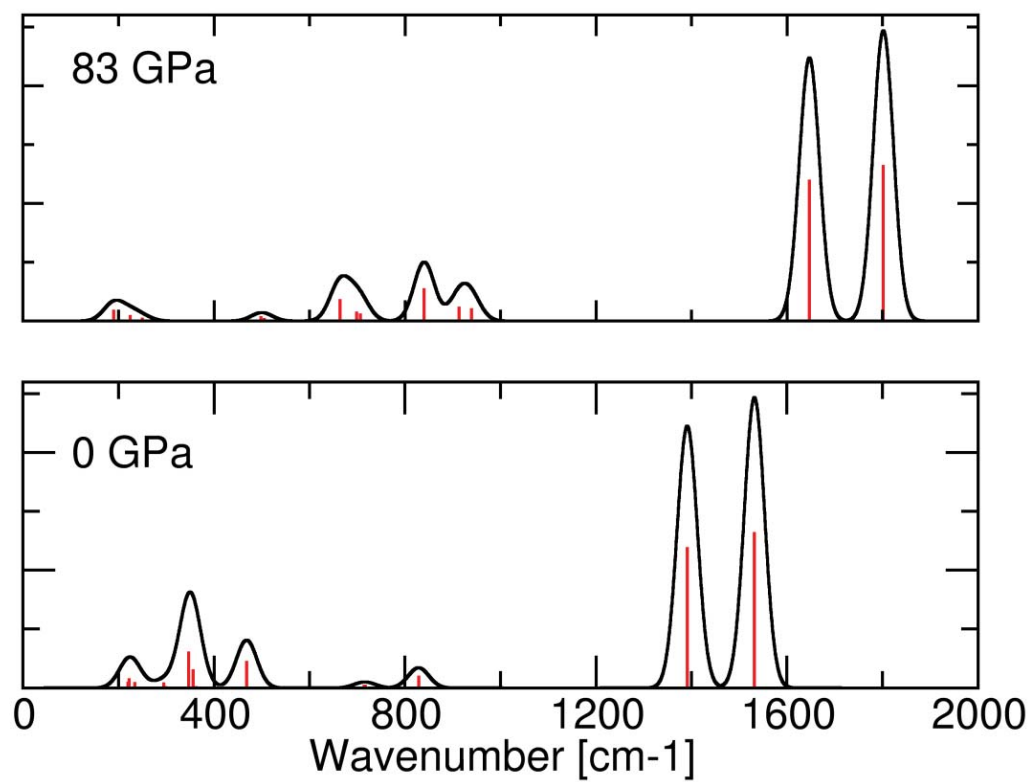

**Supplementary Figure 4:** Calculated IR spectra of magnesite at ambient pressure and 83 GPa. The LO-TO splitting is along the *y* direction.

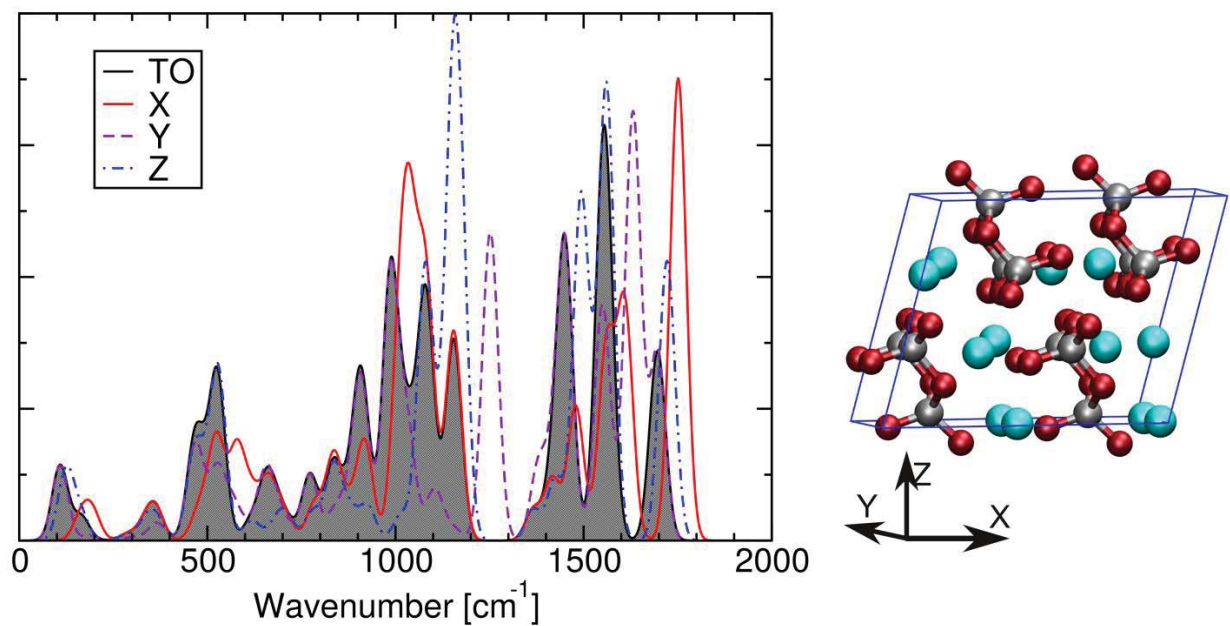

**Supplementary Figure 5:** Calculated IR spectra of the post-magnesite phase (using the experimental lattice constants). The LO-TO splitting is along *x*, *y* and *z* directions, respectively, and the pure TO mode is shown by the shading.

**Supplementary Table 1:**

Crystal structure of magnesite (space group R-3c) used for the DFT calculation

**Unit cell dimensions:**

| Ambient Conditions: | 83 GPa |
|---|---|
| a = 5.6752 Å | a = 4.83 Å |
| b = 5.6752 Å | b = 4.83 Å |
| c = 5.6752Å | c = 4.83 Å |
| alpha = 48.18° | alpha = 53.32° |
| beta = 48.18° | beta = 53.32° |
| gamma = 48.18° | gamma = 53.32° |

**Atomic coordinates:**

Ambient Conditions:

| Atom | x | y | z |
|---|---|---|---|
| Mg | 0 | 0 | 0 |
| C | 0.25 | 0.25 | 0.25 |
| O | 0.5292 | -0.0292 | 0.25 |
| O | 0.25 | 0.5292 | -0.0291 |
| O | -0.0292 | 0.25 | 0.5292 |
| Mg | 0.5 | 0.5 | 0.5 |
| C | 0.75 | 0.75 | 0.75 |
| O | 0.4708 | 1.0292 | 0.75 |
| O | 0.75 | 0.4709 | 1.0292 |
| O | 1.0292 | 0.75 | 0.4709 |

83 GPa

| Atom | x | y | z |
|---|---|---|---|
| Mg | 0.0000 | 0.0000 | 0.0000 |
| C | 0.2500 | 0.2500 | 0.2500 |
| O | 0.5390 | 0.9611 | 0.2500 |
| O | 0.2500 | 0.5390 | 0.9611 |
| O | 0.9611 | 0.2500 | 0.5390 |
| Mg | 0.5000 | 0.5000 | 0.5000 |
| C | 0.7500 | 0.7500 | 0.7500 |
| O | 0.4611 | 0.0390 | 0.7500 |
| O | 0.7500 | 0.4611 | 0.0390 |
| O | 0.0390 | 0.7500 | 0.4611 |

**Supplementary Table 2:**

Crystal structure of the post-magnesite (space group C2/m) used in the present DFT calculation

**Unit cell dimensions:**
a = 8.39 Å
b = 6.41 Å
c = 6.82 Å
alpha = 90°
beta = 105.49°
gamma = 90°

**Atomic coordinates:**

| Atom | x | y | z |
| --- | --- | --- | --- |
| C | 0.3708 | 0.3191 | 0.325 |
| C | 0.6292 | 0.6809 | 0.675 |
| C | 0.1292 | 0.8191 | 0.675 |
| C | 0.8708 | 0.1809 | 0.325 |
| C | 0.6292 | 0.3191 | 0.675 |
| C | 0.3708 | 0.6809 | 0.325 |
| C | 0.8708 | 0.8191 | 0.325 |
| C | 0.1292 | 0.1809 | 0.675 |
| C | 0.2324 | 0.5 | 0.0321 |
| C | 0.7676 | 0.5 | 0.9679 |
| C | 0.2676 | 0 | 0.9679 |
| C | 0.7324 | 0 | 0.0321 |
| Mg | 0 | 0.7536 | 0 |
| Mg | 0 | 0.2464 | 0 |
| Mg | 0.5 | 0.2536 | 0 |
| Mg | 0.5 | 0.7464 | 0 |
| Mg | 0.3276 | 0.5 | 0.692 |
| Mg | 0.6724 | 0.5 | 0.308 |
| Mg | 0.1724 | 0 | 0.308 |
| Mg | 0.8276 | 0 | 0.692 |
| Mg | 0.9331 | 0.5 | 0.6566 |
| Mg | 0.0669 | 0.5 | 0.3434 |
| Mg | 0.5669 | 0 | 0.3434 |
| Mg | 0.4331 | 0 | 0.6566 |
| O | 0.0928 | 0.5 | 0.8963 |
| O | 0.9072 | 0.5 | 0.1037 |
| O | 0.4072 | 0 | 0.1037 |
| O | 0.5928 | 0 | 0.8963 |
| O | 0.3482 | 0.1654 | 0.4286 |
| O | 0.6518 | 0.8346 | 0.5714 |
| O | 0.1518 | 0.6654 | 0.5714 |
| O | 0.8482 | 0.3346 | 0.4286 |
| O | 0.6518 | 0.1654 | 0.5714 |
| O | 0.3482 | 0.8346 | 0.4286 |
| O | 0.8482 | 0.6654 | 0.4286 |
| O | 0.1518 | 0.3346 | 0.5714 |
| O | 0.5048 | 0.3086 | 0.2677 |
| O | 0.4952 | 0.6914 | 0.7323 |
| O | 0.9952 | 0.8086 | 0.7323 |
| O | 0.0048 | 0.1914 | 0.2677 |
| O | 0.4952 | 0.3086 | 0.7323 |
| O | 0.5048 | 0.6914 | 0.2677 |
| O | 0.0048 | 0.8086 | 0.2677 |
| O | 0.9952 | 0.1914 | 0.7323 |
| O | 0.3631 | 0.5 | 0.9701 |
| O | 0.6369 | 0.5 | 0.0299 |
| O | 0.1369 | 0 | 0.0299 |
| O | 0.8631 | 0 | 0.9701 |
| O | 0.3555 | 0.5 | 0.4268 |
| O | 0.6445 | 0.5 | 0.5732 |
| O | 0.1445 | 0 | 0.5732 |
| O | 0.8555 | 0 | 0.4268 |
| O | 0.2267 | 0.3303 | 0.1572 |
| O | 0.7733 | 0.6697 | 0.8428 |
| O | 0.2733 | 0.8303 | 0.8428 |
| O | 0.7267 | 0.1697 | 0.1572 |
| O | 0.7733 | 0.3303 | 0.8428 |
| O | 0.2267 | 0.6697 | 0.1572 |
| O | 0.7267 | 0.8303 | 0.1572 |
| O | 0.2733 | 0.1697 | 0.8428 |

**Supplementary Table 3:**

Calculated and experimental IR modes of the post-magnesite, as shown in Figure 3.

| Calculated wavenumber (cm$^{-1}$) | IR intersity [(D/Å)$^2$/amu] | Experimental wavenumber (cm$^{-1}$) |
|---|---|---|
| 108.56 | 30.3446 | |
| 187.17 | 1.8279 | |
| 202.73 | 0.3101 | |
| 287.26 | 0.3745 | |
| 306.74 | 1.0823 | |
| 344.01 | 0.8428 | |
| 362.68 | 4.491 | |
| 376.57 | 2.6016 | |
| 462.71 | 24.6124 | |
| 465.23 | 0.1354 | |
| 474.97 | 14.8378 | |
| 487.47 | 0.618 | |
| 500.95 | 4.0222 | |
| 528.34 | 26.6195 | |
| 569.9 | 13.998 | |
| 606.53 | 2.1857 | |
| 631.08 | 6.4191 | |
| 658.77 | 25.2418 | 723 |
| 691.75 | 7.8076 | |
| 695.03 | 0.9758 | |
| 720.97 | 2.8591 | |
| 771.67 | 25.5318 | 805 |
| 788.83 | 1.6374 | |
| 829 | 0.4301 | |
| 834.49 | 7.1178 | |
| 852.08 | 0.9529 | |
| 852.39 | 5.7063 | |
| 879.53 | 6.7588 | |
| 886 | 4.9604 | |
| 908.17 | 59.5029 | 894 |
| 929.62 | 3.988 | |
| 987.31 | 109.1435 | 992 |
| 1031.71 | 43.2995 | |
| 1095.66 | 11.2741 | 1153 |
| 1112.3 | 11.1462 | |
| 1155.65 | 4.5695 | |
| 1252.03 | 124.4275 | 1303 |
| 1379.45 | 28.8721 | |
| 1416.96 | 22.526 | |
| 1450.63 | 117.1906 | |
| 1550.24 | 95.1927 | |
| 1631.25 | 172.6689 | |
| 1696.57 | 76.4934 | |

**Supplementary References:**

1. Haavik, C., Stolen, S., Fjellvag, H., Hanfland, M. & Hausermann, D. Equation of state of magnetite and its high-pressure modification : Thermodynamics of the Fe-O system at high pressure. *Am. Mineral.* **85,** 514–523 (2000).

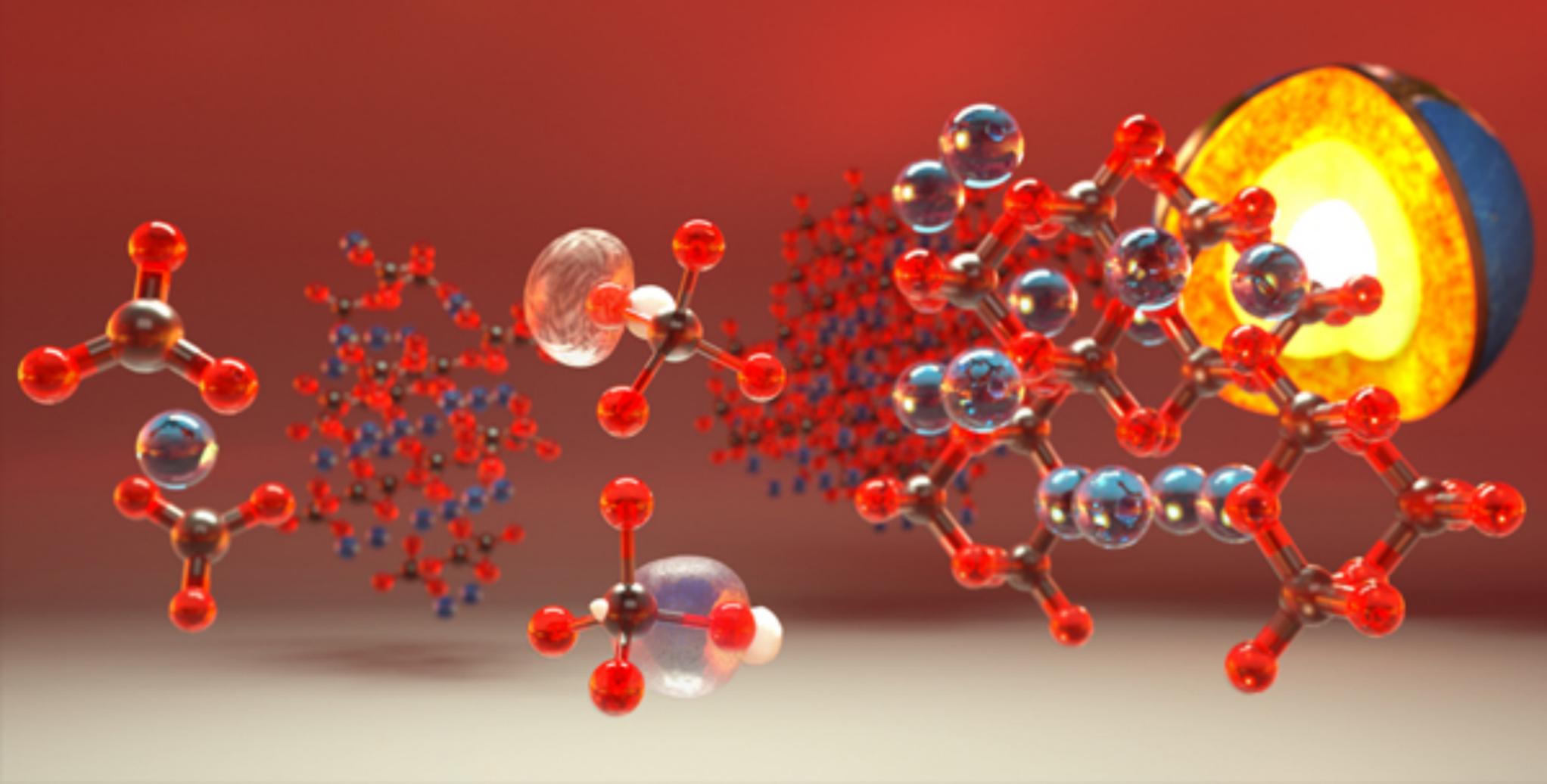